# Scaled Particle Theory for Hard Sphere Pairs. I. Mathematical Structure

Frank H. Stillinger[1], Pablo G. Debenedetti[2], and Swaroop Chatterjee[2]

[1] Department of Chemistry, Princeton University, Princeton, NJ 08544.
[2] Department of Chemical Engineering, Princeton University, Princeton, NJ 08544.

## Abstract

We develop an extension of the original Reiss-Frisch-Lebowitz scaled particle theory that can serve as a predictive method for the hard sphere pair correlation function $g(r)$. The reversible cavity creation work is analyzed both for a single spherical cavity of arbitrary size, as well as for a pair of identical such spherical cavities with variable center-to-center separation. These quantities lead directly to prediction of $g(r)$. Smooth connection conditions have been identified between the small-cavity situation where the work can be exactly and completely expressed in terms of $g(r)$, and the large-cavity regime where macroscopic properties become relevant. Closure conditions emerge which produce a nonlinear integral equation that must be satisfied by the pair correlation function. This integral equation has a structure which straightforwardly generates a solution that is a power series in density. The results of this series replicate the exact second and third virial coefficients for the hard sphere system via the contact value of the pair correlation function. The predicted fourth virial coefficient is approximately 0.6 percent lower than the known exact value. Detailed numerical analysis of the nonlinear integral equation has been deferred to the sequel (following paper).

## I. Introduction

The conceptually simple hard sphere model has served for well over a century in research efforts to understand a wide range of many-body phenomena. [1-3]. But owing to the complexity of these phenomena both within and outside the regime of thermal equilibrium, this model and its extensions to other spatial dimensions continue to be the focus of basic research inquiry. The present paper ("I") and its sequel ("II") add to this lengthy sequence of investigations, but in directions that previously have been largely unexplored.

The "scaled particle theory" represents one of the insightful strategies that have been applied to the equilibrium behavior of the hard sphere model. This approach was originated by Reiss, Frisch, and Lebowitz [4] as a statistical mechanical descriptive device for the fluid phase of the model in three dimensions, with results that agreed well with computer simulations that were available at the time [5-7]. Several refinements to the method have appeared in the literature [8,9]. Subsequent work utilizing the strategy of the scaled-particle theory has provided useful contributions to fluid mixtures [10], to other hard-particle convex shapes [11,12], to hydrophobic phenomena [13,14], and to crystals [15]. The primary objective of this and the following paper is to adapt and generalize the concepts of the established scaled particle theory to prediction of the hard-sphere pair correlation function, a basic attribute of the model's equilibrium state. At least one previous scaled-particle theory analysis was directed toward evaluating pair correlation functions [16], but its approach and implementation differed from those developed in the present pair of papers. It might also be noted in passing that an early paper by Meeron and Siegert [17] examined arrangements of multiple unit-size cavities in hard sphere systems, thus connecting to some aspects of the present work.

The following Section II briefly reviews the conventional scaled particle approach that employs a single variably-sized spherical exclusion region, and extends its scope to the case of a pair of exclusion regions. This pair is restricted to equal sizes, but with variable center-to-center separation, and thus it connects directly to the distance variation of the model's pair correlation function. The free energy associated with the cavity pair is determined in principle by molecular distribution functions of all orders for the hard sphere model, and that generates a family of size/separation geometric connection conditions for the double cavity that are discussed at length in Section III. In the limit that the two cavities become very large, their free energy possesses an asymptotic expansion, details of which are examined in Section IV. Section V assembles the preceding concepts in a way that produces a self-consistent format for calculating the hard sphere equilibrium pair correlation function. This format admits a fully self-determined density expansion for the pair correlation function and thus for the pressure virial coefficients, some results for which are presented in Section VI. The final Section VII examines a few issues raised by the approach developed in this paper. An Appendix collects eight mathematical conditions that form the basis for closure of the extended scaled particle theory, and that are available for later numerical study.

## II. Background and General Strategy

The hard-sphere many-body system is defined by the singular pair potential $u(r_{ij})$ acting between all pairs of particles $i,j$:

$$u(r_{ij}) = +\infty \qquad (0 \leq r_{ij} < \sigma)$$
$$= 0 \qquad (\sigma \leq r_{ij}) .$$
(II.1)

Here $\sigma > 0$ is the collision diameter for the particles. The number density $\rho = N/V$ in the large-system limit is geometrically constrained so as not to exceed the close-packing limit $\rho\sigma^3 = 2^{1/2}$. This upper density limit is attained in any of the crystalline close-packed structures for the repelling spheres (face-centered cubic, hexagonal close-packed, and their stacking hybrids), so we require that:

$$0 \leq \rho\sigma^3 \leq 2^{1/2}.$$
(II.2)

Within the domain of classical statistical mechanics, the equilibrium pair correlation function $g(r,\rho)$, normalized to unity in the large-$r$ limit, determines the pressure equation of state through its contact value via the hard-sphere version of the virial relation [18]:

$$\beta p = \rho + (2\pi\rho^2\sigma^3/3)g(\sigma,\rho) .$$
(II.3)

Here $\beta = 1/k_B T$ denotes the inverse temperature. Provided that $\rho\sigma^3 < 2^{1/2}$, the pair correlation function will be a bounded, continuous, and differentiable function of separation distance for $r > \sigma$.

One of the basic quantities considered in the conventional scaled particle theory [4] is the reversible isothermal work $W(\lambda,\rho)$ that must be expended in the large-system limit to expel all spherical particle centers from a spherical region with radius $\lambda$. If the equilibrium phase of the system is fluid, this quantity will automatically be independent of position, while in the crystalline region of the phase diagram, the same will be true if system-translation-permitting periodic boundary conditions are present, as we shall assume in this analysis. The Boltzmann factor associated with $W$ gives the probability $P_0(\lambda,\rho)$ that the spherical region would spontaneously be found to be empty:

$$P_0(\lambda,\rho) = \exp[-\beta W(\lambda,\rho)]$$
(II.4)

$$\equiv 1 - \sum_{n=1}^{\infty} P_n(\lambda, \rho) \ .$$

Here $P_n$ stands for the probability that the spherical region to be identified as the $\lambda$ cavity would be found spontaneously to contain exactly $n$ particle centers at the given density.

On account of the repulsions that act between hard spheres, the summation shown in the second line of Eq. (II.4) automatically terminates at a finite $n$ value that increases with $\lambda$, to be denoted by $n^*(\lambda)$. Table 1 presents information about the relevant $\lambda$ intervals for $1 \leq n^* \leq 7$. When $0 \leq \lambda \leq \sigma/2$ no more than a single sphere center can reside at any instant within the radius-$\lambda$ cavity ($n^* = 1$), leading in turn to the elementary result:

$$\beta W(\lambda, \rho) = -\ln(1 - 4\pi\rho\lambda^3/3) \qquad (0 \leq \lambda \leq \sigma/2) \ . \qquad (II.5)$$

As $\lambda$ increases beyond $\sigma/2$ further terms in the summation in Eq. (II.4) begin to contribute to $\beta W$, but in any case this quantity remains a continuous and differentiable function of $\lambda$ for all positive values of that variable. In the large-$\lambda$ asymptotic limit an appeal to macroscopic behavior indicates that $\beta W$ should be dominated by work contributions determined by the cavity volume and surface [4], so that:

$$\beta W(\lambda, \rho) \approx (4\pi\beta p/3)\lambda^3 + 4\pi\beta\gamma\lambda^2 - 8\pi\beta\gamma\delta\lambda \ . \qquad (II.6)$$

Here $\gamma(\rho)$ is a planar-surface free energy per unit area (surface tension for the fluid phase) for which the exclusion surface serves as the dividing surface, and $\delta(\rho)$ is a Tolman length [20] that arises from the curvature dependence of that surface free energy.

When $\lambda = \sigma$ the empty cavity is configurationally indistinguishable from one of the hard spheres comprising the many-particle system. At this stage one can verify that $n^*(\sigma) = 12$. The density of particle centers at the surface of the $\lambda = \sigma$ cavity is equal to $\rho g(\sigma, \rho)$, which is directly related to the pressure via Eq. (II.3). This connection leads to a differential work identity involving the $\lambda$ derivative of $W$:

$$\beta[\partial W(\sigma, \rho)/\partial\lambda] = 4\pi\rho\sigma^2 g(\sigma, \rho) \ . \qquad (II.7)$$

The conventional scaled particle theory assumes that expression (II.6) provides an adequate approximation to $\beta W$ for all $\lambda \geq \sigma/2$. Then the requirements that $\beta W$ and its first $\lambda$ derivative match that of the exact small-$\lambda$ form in Eq. (II.5), and that Eq. (II.7) be obeyed, suffice to determine the unknowns $p, \gamma$, and $\delta$ as functions of the density. The pressure equation of state obtained this way is the following [4]:

$$\frac{\beta p}{\rho} = \frac{1 + (\pi\rho\sigma^3/6) + (\pi\rho\sigma^3/6)^2}{[1 - (\pi\rho\sigma^3/6)]^3} \ . \qquad (II.8)$$

While this result constitutes an adequate description of the hard sphere model's fluid phase, it fails to capture the known first-order freezing transition [21]. Furthermore the simple form in Eq. (II.8) erroneously predicts a finite pressure at and beyond the close-packed density $\rho\sigma^3 = 2^{1/2}$.

The various intriguing concepts and connections advanced by the conventional scaled particle theory invite extensions that might improve upon its predictive ability. This paper proposes such an extension, specifically focusing on the statistical thermodynamics of double cavities. These are illustrated in Figure 1. They consist of two empty spherical regions with a common radius $\lambda$, whose centers are separated by distance $r \geq 0$. When $0 \leq r < 2\lambda$ the cavities overlap and constitute a singly-connected exclusion region. When $r > 2\lambda$ the exclusion region is disconnected. Two elementary geometric properties of the double cavity are its volume:

$$v(r,\lambda) = \pi[(4\lambda^3/3) + \lambda^2 r - (r^3/12)] \qquad (0 \leq r \leq 2\lambda)$$

(II.9)

$$= 8\pi\lambda^3/3 \qquad (2\lambda \leq r),$$

and its surface area:

$$a(r,\lambda) = 2\pi\lambda(2\lambda + r) \qquad (0 \leq r \leq 2\lambda)$$

(II.10)

$$= 8\pi\lambda^2 \qquad (2\lambda \leq r).$$

When the two portions of the double cavity overlap, the length of the circular perimeter common to the two spherical surfaces is:

$$l(r,\lambda) = \pi(4\lambda^2 - r^2)^{1/2} \qquad (0 \leq r \leq 2\lambda), \qquad (II.11)$$

and the angle at which these surfaces intersect is:

$$\theta(r,\lambda) = \pi - \arccos[1 - (r^2/2\lambda^2)]. \qquad (II.12)$$

This more elaborate scenario requires introduction of $W_2(r,\lambda,\rho)$, the reversible isothermal work required to clear particle centers from the interior of the double cavity. However it is important to note that the conventional scaled particle approach is automatically included in the double cavity extension. This is clear by virtue of the obvious identity for $r = 0$:

$$W_2(0,\lambda,\rho) \equiv W(\lambda,\rho). \qquad (II.13)$$

Furthermore, when $r$ becomes large the two disconnected parts of the cavity approach statistical independence so that:

$$\lim_{r \to \infty} W_2(r,\lambda,\rho) = 2W(\lambda,\rho). \qquad (II.14)$$

The pair correlation function $g(r,\rho)$ for the hard sphere system can be viewed as the Boltzmann factor for the reversible isothermal work necessary to bring a pair of spheres from infinity to finite distance $r \geq \sigma$. This implies that:

$$g(r,\rho) = \exp[2\beta W(\sigma,\rho) - \beta W_2(r,\sigma,\rho)] \qquad (r \geq \sigma),$$

(II.15)

$$= 0 \qquad (0 \leq r < \sigma).$$

Consequently the intended extension of the scaled particle approach has the capacity to become a predictive mechanism for the entire distance dependence of pair correlation, not just its contact value. When properly exploited in a self-consistent manner as described below, this feature generates a substantial improvement in the description of the single cavity quantity $W(\lambda,\rho)$ compared to that attained in its conventional analysis [4].

The overall strategy for calculating $g(r,\rho)$ from Eq. (II.15), to be explained in detail in following sections, can be summarized as follows:

(1) Two expressions for $\beta W(\lambda,\rho)$ are developed. One is valid for $\sigma/2 < \lambda < \sigma/3^{1/2}$, and another asymptotic, large-$\lambda$ form that is valid for $\lambda > \sigma/3^{1/2}$. The $\rho$-dependent coefficients of the latter are determined by invoking thermodynamic conditions at $\lambda = \sigma$ [Eqs. (II.3) and (II.7)], and by matching the $\sigma/2 < \lambda < \sigma/3^{1/2}$ and the $\lambda > \sigma/3^{1/2}$ expressions and their corresponding first two derivatives with respect to $\lambda$, at $\lambda = \sigma/3^{1/2}$.

(2) Two expressions for $\beta W_2(r,\lambda,\rho)$ are developed. One is valid for $n_{max}(r,\lambda) = 2$, and the second is an asymptotic, large-$\lambda$ form. Here $n_{max}(r,\lambda)$ denotes the maximum number of hard sphere centers that can be accommodated in a $(r,\lambda)$ double cavity. The unknown coefficients of the asymptotic form are obtained by matching the $n_{max} = 2$ and the large-$\lambda$ expressions and their first two derivatives with respect to $\lambda$ at $\lambda = \sigma/2$.

(3) Knowledge of $\beta W(\lambda = \sigma, \rho)$ and $\beta W_2(r, \lambda = \sigma, \rho)$ allows the calculation of $g(r,\rho)$, according to Eq. (II.15).

(4) The $\sigma/2 < \lambda < \sigma/3^{1/2}$ expression for $\beta W(\lambda,\rho)$ and the $n_{max} = 2$ expression for $\beta W_2(r,\lambda,\rho)$ require $g(r,\rho)$ as an input. This generates an iterative procedure for calculating $g(r,\rho)$, whereby the current estimate of this function is used as input to the calculation, and convergence is attained when the difference between this estimate and the result of computing the right hand side of Eq. (II.15) is smaller than some imposed convergence criterion.

## III. Geometric Connection Conditions

Equation (II.4) has a direct extension to the double cavity:

$$P_{0,2}(r,\lambda,\rho) = \exp[-\beta W_2(r,\lambda,\rho)]$$

(III.1)

$$\equiv 1 - \sum_{n=1}^{\infty} P_{n,2}(r,\lambda,\rho) \ .$$

In analogy to the single cavity case, $P_{n,2}$ stands for the probability that the interior of a region identical in shape to that of the double cavity, placed at random in the hard sphere system, would happen to contain the centers of exactly $n$ spherical particles. Also as before, the sum shown in the last equation necessarily truncates at some finite upper limit on $n$, now dependent on the values of both $r$ and $\lambda$, which will be denoted by $n_{max}(r,\lambda)$ as introduced earlier.

In the preceding background Section II we have used $\sigma$-explicit expressions for consistency with the literature [4]. The natural distance unit choice for the hard sphere model is the collision diameter, so in what follows we set $\sigma = 1$, and hence have the density restriction $0 \leq \rho \leq 2^{1/2}$. As a result of this convention, it becomes unnecessary for $\sigma$ to appear explicitly in any of the following mathematical expressions.

When $r > 2\lambda + 1$ any spheres occupying one of the disconnected halves of the double cavity cannot geometrically interfere with spheres in the other half. Consequently one has:

$$n_{max}(r,\lambda) = 2n^*(\lambda) \qquad (r > 2\lambda + 1) \ . \qquad \text{(III.2)}$$

For smaller distances $r$, interferences can come into play, and can have the effect of reducing $n_{max}$. This interference reaches its greatest extent at $r = 0$, for which obviously:

$$n_{max}(0,\lambda) \equiv n^*(\lambda) \ . \qquad \text{(III.3)}$$

In the infinite system limit that is the object of attention in this study, one can introduce a full family of symmetric particle correlation functions $g^{(n)}(\mathbf{r}_1...\mathbf{r}_n,\rho)$ for sets of $n = 1,2,3,...$ spheres at positions $\mathbf{r}_1...\mathbf{r}_n$. The non-overlap property of hard spheres of course implies that any $g^{(n)}$ vanishes if one or more of its pair distances $r_{ij} < 1$. On account of the free translation permitted by periodic boundary conditions, one has $g^{(1)} \equiv 1$ even in the crystal phase, and the higher-order $g^{(n)}$ depend only on relative, not absolute, positions. In the fluid phase all $g^{(n)}$ approach unity in the limit that all of its pair distances diverge. The function $g(r,\rho)$ appearing in Eq. (II.3) is identical to $g^{(2)}(\mathbf{r},\rho)$ in the equilibrium range of fluid densities, and if necessary it can be construed as an orientational average of $g^{(2)}(\mathbf{r},\rho)$ in the crystal density range. It is possible to convert expression (III.1) into an equivalent alternating series whose terms involve integrals of the $g^{(n)}$ over the interior of the double cavity [4,22]:

$$\exp[-\beta W_2(r,\lambda,\rho)] = 1 + \sum_{n=1}^{n_{max}} \frac{(-1)^n \rho^n}{n!} \int_v d\mathbf{r}_1 ... \int_v d\mathbf{r}_n g^{(n)}(\mathbf{r}_1...\mathbf{r}_n,\rho) \ . \qquad \text{(III.4)}$$

In particular, the individual particle integrals each span the full interior of the double cavity with volume $v(r,\lambda)$, even when it comprises two disconnected portions.

In order for $n_{max}(r,\lambda)$ to equal its minimum value unity, it is necessary for the most remote pair of distances in the double cavity region to be less than the distance of closest approach for spheres. This is satisfied if:

$$r + 2\lambda < 1 \qquad (n_{max} = 1) \ . \qquad \text{(III.5)}$$

The correspondingly truncated form of the series in Eq. (III.4) leads to:

$$\beta W_2(r,\lambda,\rho) = -\ln[1 - \rho v(r,\lambda)] \qquad (r + 2\lambda \leq 1) \ , \qquad \text{(III.6)}$$

an obvious generalization of earlier Eq. (II.5).

As Eq. (III.2) indicates, when $\lambda < 1/2$ and $r$ is sufficiently large, each of the disconnected portions of the double cavity can accommodate a single sphere center, i.e. $n_{max} = 2$. This upper limit on $\lambda$ for the $n_{max} = 2$ region remains in force until $r$ declines into the interval:

$$0 \leq r < (3^{1/2} - 1)/2 \cong 0.366025 \ . \qquad \text{(III.7)}$$

In this restricted $r$ range the mutual interference between spheres permits $\lambda$ to increase somewhat above 1/2 while still excluding a third sphere from invading the interior of the double cavity. The upper limit for this $\lambda$ increase is established by considering the extremal geometry illustrated in Figure 2, in which the centers of three spheres can simultaneously fit on the boundary of the double cavity. Any increase in either

$r$ or $\lambda$ (or both) would permit these three spheres some extent of independent motion inside the double cavity. Straightforward algebra leads to the following relation at the geometry shown in Figure 2:

$$r^2 + 2(r+\lambda)\{\lambda + [\lambda^2 - (1/4)]^{1/2}\} = 1 \quad . \qquad (\text{III.8})$$

This can be transformed into an explicit expression for the upper $\lambda$ limit when $r$ is in the interval (III.7). Consequently the $r$-dependent upper limit $\lambda_2(r)$ for the $n_{max} = 2$ region in the $r, \lambda$ positive quadrant has the following form:

$$\lambda_2(r) = \frac{1}{2}\left(\frac{3^{1/2}}{2} - r\right) + \frac{1}{8}\left(\frac{3^{1/2}}{2} - r\right)^{-1} \qquad [0 \le r < (3^{1/2} - 1)/2] \; ,$$

(III.9)

$$= 1/2 \qquad\qquad [(3^{1/2} - 1)/2 \le r] \; .$$

Regions corresponding to odd integer values of $n_{max}$ are necessarily confined to relatively small $r$ on account of Eq. (III.2). The case $n_{max} = 3$ is no exception. Its upper boundary $\lambda_3(r)$ is determined by a tetrahedral arrangement of the centers of four mutually-contacting spheres on the surface of the double cavity, with two of the four on each of its two $\lambda$-sphere portions. This extremal geometry leads to the following expression for the $n_{max} = 3$ upper boundary:

$$\lambda_3(r) = (1/2)[1 + (r - 2^{-1/2})^2]^{1/2} \qquad (0 \le r < 2^{-1/2}) \; . \qquad (\text{III.10})$$

No portion of the $n_{max} = 3$ region exists for $2^{-1/2} < r$. Figure 3 shows the regions for $n_{max} = 1, 2, 3$ and $\ge 4$ in the $r, \lambda$ positive quadrant.

Provided that $\lambda < 1/2$ for any $r \ge 0$, it is possible to express the $n = 2$ term in the series (III.4) as a simple quadrature involving $g(r, \rho)$. This results from the fact that the integral can be expressed in a convolution form, a reduction that is not applicable within that small portion of the $n_{max} = 2$ strip near the origin that rises above $\lambda = 1/2$. With this restriction one has:

$$\beta W_2(r, \lambda, \rho) = -\ln[1 - \rho v(r, \lambda) + I(r, \lambda, \rho)] \; , \qquad (n_{max} = 2, \; \lambda < 1/2) \qquad (\text{III.11})$$

where the explicit form of the pair correlation contribution is the following:

$$I(r, \lambda, \rho) = (2\pi^2 \rho^2 / r)\int_{\max(r-2\lambda, 1)}^{r+2\lambda} k(r - r_1, \lambda) r_1 g(r_1, \rho) dr_1 \; ,$$

(III.12)

$$k(u, \lambda) = (8/15)\lambda^5 - (2/3)\lambda^3 u^2 + (1/3)\lambda^2 |u|^3 - (1/60)|u|^5 \; .$$

As Eq. (II.14) has already indicated, the large-$r$ limit for the double cavity reduces to twice that for the single cavity with the same $\lambda$ value. In this single-cavity circumstance series (III.4) with $r = 0$ terminates at $n = 2$ over the interval $1/2 \le \lambda \le 3^{-1/2}$. Once again the $n = 2$ integral has a convolution form and may be reduced to a simple quadrature [though the result cannot be immediately derived from Eq. (III.12) above]. One finds:

$$\beta W(\lambda, \rho) = -\ln[1 - (4\pi\rho\lambda^3/3) + H(\lambda, \rho)] \qquad (1/2 \le \lambda \le 3^{-1/2}) \; ,$$

$$H(\lambda,\rho) = 2\pi^2 \rho^2 \int_1^{2\lambda} [(4/3)\lambda^3 r_1^2 - \lambda^2 r_1^3 + (1/12) r_1^5] g(r_1,\rho) dr_1 \ . \qquad \text{(III.13)}$$

Suppose $\lambda$ were to exceed $1/2$ only slightly in this last expression, specifically:

$$\lambda = (1+\varepsilon)/2 \qquad \text{(III.14)}$$

where $\varepsilon$ is very small. Then the range of $r_1$ integration in $H(\lambda,\rho)$ would be sufficiently small that in the integrand one could set $g(r_1,\rho) \approx g(1,\rho)$ to leading order. Explicit evaluation of the resulting elementary integral then leads to the conclusion:

$$\beta W(\lambda,\rho) = -\ln\{1 - (4\pi\rho\lambda^3/3) + [\pi^2 \rho^2 g(1,\rho)/6]\varepsilon^3 + O(\varepsilon^4)\} \ . \qquad \text{(III.15)}$$

Consequently the sudden inclusion of sphere pairs inside an expanding single-sphere cavity only causes a discontinuity in third and higher derivatives of $\beta W(\lambda,\rho)$ with respect to $\lambda$.

A similar conclusion applies to continuity of $\beta W_2(r,\lambda,\rho)$ and its first two $\lambda$ derivatives as $\lambda$ begins to exceed $1/2$. Although $g^{(3)}$ and $g^{(4)}$ formally appear as integrand factors in the $n_{\max} = 4$ region, these functions are finite and do not basically influence the geometry involved in the possible insertion of sphere pairs in each separate portion of the double cavity when $r \geq 2\lambda + 1$. When $r$ is small enough to permit interferences to occur ($r < 2\lambda + 1$), the available contributing configurations will be somewhat restricted. However no mechanism is present to generate any terms of lower order in $\varepsilon$ than $O(\varepsilon^3)$, so the analog to the preceding Eq. (III.15) is:

$$\beta W_2(r,\lambda,\rho) = -\ln[1 - \rho v(r,\lambda) + O(\varepsilon^3)] \qquad [\lambda = (1+\varepsilon)/2] \ . \qquad \text{(III.16)}$$

It is also useful to consider the effect on the single cavity function $\beta W(\lambda,\rho) \equiv \beta W_2(0,\lambda,\rho)$ as the size parameter $\lambda$ begins to exceed $3^{-1/2}$, thus just allowing invasion by a sphere triplet. Figure 4 provides a graphical guide for the geometry of these three particles (1, 2, and 3). The particle non-overlap constraints permit the center of particle 1 to move on a sphere of radius $\approx \lambda$, but radially within the cavity only by a small magnitude of order $\approx (\lambda - 3^{-1/2})$. With particle 1 at any of its possible locations, particle 2 can swing around a circle with radius $\approx 1/2$, but its other two degrees of freedom are each restricted to $\approx (\lambda - 3^{-1/2})$, essentially the same small parameter. With particles 1 and 2 fixed at any of their possible locations, particle 3 is restricted in all three of its degrees of motion freedom each to $\approx (\lambda - 3^{-1/2})$. By accounting for all of the small factors identified by this informal analysis, one concludes that as $\lambda$ begins to exceed $3^{-1/2}$ the corresponding contribution from the $n = 3$ term in series (III.4) for $r = 0$ is dominated by the factor $(\lambda - 3^{-1/2})^6$. This implies that only the sixth and higher $\lambda$ derivatives of $\beta W(\lambda,\rho)$ will exhibit discontinuities at $\lambda = 3^{-1/2}$. Furthermore, an extension of this type of argument suggests that even higher order derivatives are those first manifesting discontinuities when increasing $\lambda$ crosses the boundaries into the $n^* = 4, 5,\ldots$ regions, as identified in Table 1.

## IV. Asymptotic Properties

As will become clear in the following, the present double-cavity formalism offers a wider array of mathematical interconnections between its various functions than are present in the conventional scaled particle theory. Consequently it will prove to be possible to represent the single cavity work quantity $\beta W$ by a more elaborate asymptotic form than shown earlier in Eq. (II.6). Specifically we shall utilize a five-term representation:

$$\beta W(\lambda, \rho) \cong J(\rho)\lambda^3 + K(\rho)\lambda^2 + L(\rho)\lambda + M(\rho) + N(\rho)/\lambda \ . \qquad (\text{IV.1})$$

The first three coefficients $J, K$, and $L$ will necessarily have the same interpretation as before in terms of the macroscopic properties pressure, surface tension, and surface tension curvature dependence. All five coefficients in Eq. (IV.1) require self-consistent determination in the present extension.

With suitable modifications, the double cavity quantity $\beta W_2(r, \lambda, \rho)$ should also be amenable to representation by an asymptotic form analogous to that shown in Eq. (IV.1) for the single cavity function $\beta W(\lambda, \rho)$. In the large cavity regime it is again possible to identify the two leading terms as arising from macroscopic pressure-volume and area-surface tension work terms. Therefore in leading orders in $\lambda$ one can assume:

$$\beta W_2(r, \lambda, \rho) = \beta p v(r, \lambda) + \beta \gamma a(r, \lambda) + .... \ , \qquad (\text{IV.2})$$

where in principle $p$ and $\gamma$ are the same functions of density that determine $J(\rho)$ and $K(\rho)$ in Eq. (IV.1). Equation (IV.2) should be appropriate regardless of whether the two portions of the double cavity overlap or are disconnected. Note that a succeeding term of order $\lambda$ involving simply the curvature dependence of surface tension, following those shown in Eq. (IV.2), has not been invoked. This is an important observation arising from the fact that other contributions of the same linear order in $\lambda$ will begin to appear. These can be identified as due (a) to a line tension associated with the circular perimeter $l(r, \lambda)$ when $r < 2\lambda$, and (b) to interferences between the short-range correlations induced in the sphere system around each half of the double cavity.

The disconnection property of double cavities introduces an aspect not present in the conventional scaled particle theory. This feature is expected to create singularities in the functions of interest such as $\beta W_2$ at the point of disconnection ($r = 2\lambda$). For that reason it is helpful to introduce a simple linear transformation to a pair of oblique variables:

$$s = r - 2\lambda \ ,$$
$$t = 2\lambda - 1 \ , \qquad (\text{IV.3})$$

or equivalently:

$$r = s + t + 1 \ ,$$
$$\lambda = (t+1)/2 \ . \qquad (\text{IV.4})$$

Here $s$ measures the positive distance between nearest points of the two $\lambda$-spheres when they do not overlap; but when they do overlap $s$ is negative and has a magnitude equal to the distance between the interpenetrating surfaces measured along the line of centers. The other variable $t$ is just a shifted and rescaled cavity-sphere size variable. Figure 5 graphically illustrates the $s = 0$ cavity-contact line and its intersection with the $t = 1$ ($\lambda = 1$) half-line that is the locus of points involved in evaluation of the pair correlation function $g(r, \rho)$. Other constant-$s$ loci are a family of lines parallel to the one shown in Fig. 5. By holding $s \neq 0$ constant, increasing $\lambda(t)$ from small to large values does not encounter the connection/disconnection singularity, and the function and derivative continuity conditions previously stated in terms of $\lambda$ at constant $r$ can be directly transferred to variations with respect to $t$ at constant $s$.

In the large-$\lambda$ asymptotic limit, each of the geometric characteristics v, a, and $l$ of the double cavity, Eqs. (II.9-11), can be expanded in descending integer powers of $\lambda$. Therefore it is a reasonable assumption that at constant $s$ the double cavity work quantity $\beta W_2$ has an asymptotic expansion that is consistent with, but extends, the leading terms shown in Eq. (IV.1) above. For the remainder of this paper, we shall suppose that a five-term representation is possible at any chosen density $\rho$, analogous to that for $\beta W$:

$$\beta W_2(r,\lambda,\rho) \cong J(\rho)\left[\frac{v(r,\lambda)}{v(0,\lambda)}\right]\lambda^3 + K(\rho)\left[\frac{a(r,\lambda)}{a(0,\lambda)}\right]\lambda^2 + X(s,\rho)\lambda + Y(s,\rho) + \left[\frac{Z(s,\rho)}{\lambda}\right] . \qquad (\text{IV.5})$$

Here $r$ and $\lambda$ in the right member can be expressed in terms of their $s$ and $t$ equivalents. The cavity interference effects mentioned above cause the new coefficient functions $X$, $Y$, and $Z$, dependent on separation variable $s$, to appear in this expression in place of $L$, $M$, and $N$.

## V. Self-Consistency Conditions

The asymptotic expressions (IV.1) and (IV.5) for $\beta W$ and $\beta W_2$ are smooth functions of the cavity size parameter $\lambda$, and thus are intrinsically incapable of representing the singularities that occur across each locus at which $n^*$ or $n_{max}$ change discontinuously. However it has been stressed above in Section III that these singularities affect only high-order derivatives, particularly when large $n^*$ or $n_{max}$ are involved. Likewise, weak singularities in the $g^{(n)}$ themselves [23], appearing as integrand factors in series (III.4), should only influence high-order derivatives of $\beta W$ and $\beta W_2$. Consequently it is a sensible approximation to suppose that the forms of asymptotic expressions (IV.1) and (IV.5) can be extended to relatively small $\lambda$ with appropriate choices for the coefficients. In fact we shall assume that Eq. (IV.1) for $\beta W(\lambda,\rho)$ is valid for:

$$3^{-1/2} \leq \lambda < +\infty , \qquad (\text{V.1})$$

which covers the entire range over which $n^* \geq 3$. In the case of the double cavity, the corresponding assumption will be that asymptotic expression (IV.5) adequately covers the following $\lambda$ range:

$$1/2 \leq \lambda < +\infty , \qquad (\text{V.2})$$

which by Eq. (IV.3) is equivalent to:

$$0 \leq t < +\infty . \qquad (\text{V.3})$$

The fact that different $\lambda$ lower limits are to be used for these two cases does not in itself constitute an inconsistency, because distinct sets of paths in the $r,\lambda$ positive quadrant will be involved. In order to evaluate the hard-sphere pair correlation function $g(r,\rho)$ it will be necessary to determine $\beta W$ along the vertical ($r=0$) axis, and to determine $\beta W_2$ along inclined linear paths of constant $s \geq -1$ that intersect the $r \geq 1$, $\lambda = 1$ half-line, as indicated in Fig. 5.

If a complete strategy is to be constructed, it must contain enough relations to determine the eight unknown quantities $J$, $K$, $L$, $M$, $N$, $X(s)$, $Y(s)$, and $Z(s)$ as functions of $\rho$. Suppose initially $g(r,\rho)$ has been specified for the density of interest, if only in approximate form. This immediately allows evaluation of the pressure [Eq. (II.3)] and hence evaluation of $J \, [= 4\pi\beta p/3]$. Using this same input $g(r,\rho)$, Eq.

(III.13) can be used to provide values at $\lambda = 3^{-1/2}$ for $\beta W(\lambda, \rho)$ and its first two $\lambda$ derivatives, and because each of these must be continuous at that point these values constitute three constraints on $K$, $L$, $M$, and $N$ that are the remaining four coefficients in the $\beta W(\lambda, \rho)$ asymptotic form, Eq. (IV.1). The $\lambda = 1$ identity, Eq. (II.7), is the final condition needed to fix the five coefficients $J$, ..., $N$ to be consistent with the given, possibly approximate, input $g(r, \rho)$.

The requirements of smooth connection across the $\lambda = 1/2$ line suffice to determine the remaining functions $X(s, \rho)$, $Y(s, \rho)$, and $Z(s, \rho)$. To implement these conditions in the most useful manner, the two forms of $\beta W_2$ [Eqs. (III.11) and (IV.5)] are expressed in terms of variable $t$, and then continuity of this function as well as of its first and second derivatives with respect to $t$ at $t = 0$ are imposed.

Explicit mathematical formulas for the eight conditions just described have been collected in the Appendix. Applying these formulas as evaluated with the given estimate for $g(r, \rho)$ leads to corresponding predictions for $\beta W(\lambda = 1, \rho)$ and $\beta W_2(r, \lambda = 1, \rho)$. Equation (II.15) then yields a next-stage estimate for $g(r, \rho)$. Of course the objective is self-consistency, with identical $g(r, \rho)$'s appearing both as input and output, and when that is attained at the chosen density $\rho$ it constitutes a reportable prediction of this extended version of scaled particle theory.

## VI. Density Expansion

The pair correlation functions for classical models with uncharged particles have convergent power series in density that describe those functions in the low to moderate density regime [24]. In particular this is true for the hard sphere model under consideration in this paper, so we write the formal series:

$$g(r, \rho) = \sum_{j=0}^{\infty} \rho^j g_j(r) , \qquad \text{(VI.1)}$$

where of course $g$ and all of the $g_j$ vanish identically for $r < 1$. The density-independent leading term is just the pair Boltzmann factor for the hard sphere potential, i.e. the unit step function $U$ centered at the collision diameter:

$$g_0(r) = \exp[-\beta u(r)]$$
$$\equiv U(r - 1) . \qquad \text{(VI.2)}$$

Exact evaluation of the succeeding $g_j(r)$ would involve Mayer cluster integrals whose numbers and topological complexities rise very rapidly with order $j$ [24, 25]. The $j \geq 1$ terms in Eq. (VI.1) for hard spheres vanish identically when $r \geq j + 1$. The present extended version of the scaled particle theory in principle offers an alternative method to generate the density expansion (VI.1), albeit in approximate form. The purpose of this Section VI is to demonstrate that the theory as described above indeed has the capacity to produce such a density expansion. To be able to do so should be regarded as a necessary condition that the theory must meet. A failure to meet that condition would have to be regarded as a fatal flaw invalidating the approach.

Note first that the zero-order pair function (VI.2), when inserted into the virial equation of state expression (II.3) yields the pressure correct through $O(\rho^2)$. More generally, if $g_0(r)...g_l(r)$ are known, inserting those results into the virial equation predicts pressure through $O(\rho^{l+2})$. This is also equivalent to specifying the single-cavity function $J(\rho)$ through $O(\rho^{l+2})$. When this result for $J$ is inserted in Eqs.

(A.2)-(A.5) one obtains four independent linear equations for the remaining four single-cavity unknowns $K$, $L$, $M$, and $N$, with constant coefficients, and inhomogeneous terms valid through $O(\rho^{l+1})$. Solving this equation set thus determines each of $K$, $L$, $M$, and $N$ through $O(\rho^{l+1})$.

Similar remarks apply to the three connection conditions (A.6) for the pair cavity. This insertion of $g(r,\rho)$ through $O(\rho^l)$, along with the single-cavity quantities just determined through $O(\rho^{l+1})$, suffice to determine $X(s,\rho)$, $Y(s,\rho)$, and $Z(s,\rho)$ through $O(\rho^{l+1})$. This means that both $\beta W(\lambda,\rho)$ and $\beta W_2(r,\lambda,\rho)$ can be evaluated through $O(\rho^{l+1})$. Equation (II.15) subsequently provides $g_{l+1}(r)$, thus completing a stage of the density expansion. Repetition of this procedure in principle continues to increment $l$ and so can extend the series in Eq. (VI.1) to an arbitrary number of terms.

For purposes of concrete illustration we now carry out this strategy in explicit form for the first few orders. The starting point simply utilizes the ideal gas pressure, the leading term in the right member of Eq. (II.3), namely $\beta p = \rho + O(\rho^2)$. It is then straightforward to show that in this density order, $J(\rho)$ is the only non-vanishing member of the set of eight unknowns $J(\rho), L(\rho),..., Z(s,\rho)$:

$$J(\rho) = (4\pi/3)\rho + O(\rho^2) \ . \tag{VI.3}$$

Subsequently evaluating $\beta W(\lambda,\rho)$ and $\beta W_2(r,\lambda,\rho)$ at $\lambda = 1$, and inserting the results in Eq. (II.15) yields the following:

$$g_1(r) = \pi\left(\frac{4}{3} - r + \frac{r^3}{12}\right) \qquad (1 \leq r \leq 2) \ ,$$

$$= 0 \qquad (2 \leq r) \ . \tag{VI.4}$$

This is the exact result in linear order in density, well-known from the Mayer cluster expansion [25]. When combined with $g_0(r)$, Eq. (VI.2), and placed in the virial pressure expression Eq. (II.3), this correctly reproduces the equation of state through the third virial coefficient:

$$\beta p = \rho + (2\pi/3)\rho^2 + (5\pi^2/18)\rho^3 + O(\rho^4) \ . \tag{VI.5}$$

At the next level of iteration, only the first two terms in Eq. (VI.5) for the pressure are required, in order to specify $J(\rho)$ to second order:

$$J(\rho) = (4\pi/3)\rho + (8\pi^2/9)\rho^2 + O(\rho^3) \ . \tag{VI.6}$$

This, and the pair correlation function through linear order in density, may then be inserted into each of the four connection condition Eqs. (A.2)-(A,5) to evaluate their inhomogeneous terms consistently through $O(\rho^2)$. The results are the following:

$$2K + L - N = -\pi^2\rho^2 + O(\rho^3) \ ; \tag{VI.7}$$

$$3^{1/2}K + 3L + 3^{3/2}M + 9N = -\frac{5 \cdot 3^{1/2}\pi^2\rho^2}{12} + O(\rho^3) \ ; \tag{VI.8}$$

$$2K + 3^{1/2}L - 3^{3/2}N = -\pi^2\rho^2 + O(\rho^3) \ ; \qquad\qquad \text{(VI.9)}$$

$$K + 3^{3/2}N = -\pi^2\rho^2/2 + O(\rho^3) \ . \qquad\qquad \text{(VI.10)}$$

These last four equations are easily solved through the requisite density order:

$$K = -\pi^2\rho^2/2 + O(\rho^3) \ ,$$

$$L = 0 + O(\rho^3) \ ,$$

$$M = \pi^2\rho^2/36 + O(\rho^3) \ , \qquad\qquad \text{(VI.11)}$$

$$N = 0 + O(\rho^3) \ .$$

The results shown here for $K$ and $L$ agree in this density order with their analogs in the conventional scaled particle theory, but $M$ and $N$ have no precedents in that earlier version with which to compare [4].

The next step requires setting up and solving the three equations (A.6) for the unknowns $X(s,\rho)$, $Y(s,\rho)$, and $Z(s,\rho)$ through the same second order in density. By confining attention just to a result for $g_2(1)$, which is sufficient to predict the fourth virial coefficient, we can limit the analysis to these three unknowns just at $s \to +\infty$ and at $s = -1$. For the first of these two cases, starting with Eq. (III.11), expanding in density through second order, and applying the connection conditions leads to the following:

$$X(+\infty,\rho) = 0 + O(\rho^3) \ ,$$

$$Y(+\infty,\rho) = \frac{\pi^2\rho^2}{18} + O(\rho^3) \ , \qquad\qquad \text{(VI.12)}$$

$$Z(+\infty,\rho) = 0 + O(\rho^3) \ .$$

Then using these results for subsequent evaluation of $\beta W_2$ for this infinite-separation case yields the following work quantity for insertion of two independent unit spheres:

$$\beta W_2(s = +\infty, t = 1) = \frac{8\pi\rho}{3} + \frac{5\pi^2\rho^2}{6} + O(\rho^3) \ . \qquad\qquad \text{(VI.13)}$$

The same result emerges from the alternative route that evaluates $\beta W(\lambda = 1)$ directly from the single-cavity analysis and multiplies the result by 2, as indicated in Eq. (II.14).

For the purpose of evaluating $g(1,\rho)$ it is necessary to obtain $\beta W_2(s = -1, t = 1)$ through the same quadratic order in density. The corresponding computation that also starts with Eq. (III.11) is quite tedious, so we skip most details. This procedure yields the following intermediate results:

$$X(-1,\rho) = \frac{\pi^2\rho^2}{16} + O(\rho^3) \ ,$$

$$Y(-1,\rho) = -\frac{5\pi^2\rho^2}{144} + O(\rho^3) ,  \qquad (VI.14)$$

$$Z(-1,\rho) = \frac{\pi^2\rho^2}{64} + O(\rho^3) .$$

When these are combined with J and K through the same order, one is able to establish that:

$$\beta W_2(s=-1,t=1) = \frac{9\pi\rho}{4} + \frac{457\pi^2\rho^2}{576} + O(\rho^3) . \qquad (VI.15)$$

The contact pair correlation function follows from Eq. (II.15), which requires the results in Eqs. (VI.13) and (VI.15):

$$g(1,\rho) = 1 + \frac{5\pi\rho}{12} + \frac{73\pi^2\rho^2}{576} + O(\rho^3) . \qquad (VI.16)$$

Finally the virial pressure equation of state, Eq. (II.3), can be exhibited explicitly through fourth order in density:

$$\beta p = \rho + \frac{2\pi}{3}\rho^2 + \frac{5\pi^2}{18}\rho^3 + \frac{73\pi^3}{864}\rho^4 + O(\rho^5) . \qquad (VI.17)$$

The last explicitly shown coefficient in Eq. (VI.17) is the fourth virial coefficient $B_4$ as predicted by the present extended scaled particle theory (ESPT):

$$B_4 \cong \frac{73\pi^3}{864} \qquad (ESPT) . \qquad (VI.18)$$

This is close, but not precisely equal, to the known exact value [25]:

$$B_4 = 2.6362.... \cong \frac{73.459\pi^3}{864} \qquad (Exact) . \qquad (VI.19)$$

To put this comparison in a useful context, we also note that the fourth virial coefficient estimate that emerges from the equation of state (II.8) produced by the conventional scaled particle theory (SPT) [4] is the following:

$$B_4 \cong \frac{19\pi^3}{216} \equiv \frac{76\pi^3}{864} \qquad (SPT) . \qquad (VI.20)$$

It is also worth mentioning that the popular Carnahan-Starling (C-S) approximate equation of state for the hard sphere fluid [26]:

$$\frac{\beta p}{\rho} = \frac{1 + \frac{\pi\rho}{6} + \left(\frac{\pi\rho}{6}\right)^2 - \left(\frac{\pi\rho}{6}\right)^3}{\left(1 - \frac{\pi\rho}{6}\right)^3} \tag{VI.21}$$

makes the prediction:

$$B_4 \cong \frac{\pi^3}{12} \equiv \frac{72\pi^3}{864} \qquad \text{(C-S)} \ . \tag{VI.22}$$

## VII. Discussion

The original scaled particle theory created by Reiss, Frisch, and Lebowitz focused on the creation work for a single spherical cavity with arbitrary radius in the single-component hard sphere system [4]. In addition to generating an equation of state for the fluid phase of that hard sphere system their approach automatically had the capacity to estimate the chemical potential for a dissolved spherical solute of any size, in the infinite-dilution limit. The extension presented in this and the following paper adapts concepts of that original work to the case of a double cavity composed of two spherical exclusion regions with arbitrary but equal sizes, with variable center-to-center separation. The underlying motivation was the desire to predict the distance and density dependent pair correlation function for hard spheres, $g(r, \rho)$. This extension includes description of the single cavity quantity of the original theory, but embeds that description in a somewhat more elaborate formalism. In analogy with the previous situation, this double-cavity approach in principle allows estimation of the infinite-dilution-limit chemical potential for diatomic species dissolved in the hard sphere solvent.

A necessary test that needs to be passed by any proposed approach to predicting $g(r, \rho)$ is that it can generate a well-defined density expansion for this function, and for the hard-sphere virial series. Section VI above demonstrates that such a density expansion emerges from the present strategy. In particular the second and third virial coefficients are reproduced exactly. The implied fourth virial coefficient exhibits a small deviation from the known exact value, but that deviation is substantially less than those produced either by the original scaled particle theory or the Carnahan-Starling equation of state. These comparisons indicate that the present extension of the scaled particle theory has intrinsic merit.

It needs to be stressed that the exploratory analysis offered in this paper has not invoked the full set of thermodynamic conditions that are available. Specifically, the spatial integral of the pair correlation function leads to the isothermal compressibility, whose inverse yields the pressure equation of state upon integration with respect to density. Also, the single cavity quantity $W(\lambda = 1, \rho)$ represents the non-ideal part of the hard sphere chemical potential, that also must be consistent with the virial equation of state at all densities. This unused information, if incorporated, might well serve to modify the strategy developed above to produce a more accurate theory.

The five-term asymptotic expansions, Eqs. (IV.1) and (IV.5) that were introduced for the single and double cavities respectively, have been assumed valid and used as such down to small cavity sizes. This presumption deserves careful scrutiny in the future. It is possible that the nature of local sphere arrangements, particularly at high densities, requires a different and more flexible format. In particular an accurate account of the first-order fluid-crystal transition displayed by the hard sphere system might demand such a modification. It also should be stressed that nothing is explicitly present in the double cavity formalism developed here that would assure divergence of the pressure at the close-packed density $\rho = 2^{1/2}$. The unused exact conditions mentioned in the preceding paragraph might be important components of a variant of the double-cavity scaled particle theory that successfully predicts these characteristics of the hard sphere system.

## Appendix

The purpose of this Appendix is to collect the self-consistency and connection conditions that are needed to determine $\beta W(\lambda)$ and $\beta W_2(r,\lambda)$. These conditions must suffice to determine the eight quantities $J$, $K$, $L$, $M$, $N$, $X(s)$, $Y(s)$, and $Z(s)$ as functions of density $\rho$. They will be stated in a fashion that assumes a given input estimate is available for the pair correlation function $g(r)$. For notational economy here and in the following, we suppress explicit appearance of $\rho$ as a variable in the various functions to be determined.

The first condition connects $J$ to the contact value of the pair correlation function, utilizing Eqs. (II.3), (II.6), and (IV.1):

$$J = \frac{4\pi\rho}{3} + \frac{8\pi^2\rho^2}{9} g(1) \quad . \tag{A.1}$$

A second condition emerges from Eq. (II.7), relating to incremental single-cavity expansion work at $\lambda = 1$:

$$3J + 2K + L - N = 4\pi\rho g(1) \quad . \tag{A.2}$$

The continuity of $\beta W$, of $\beta\partial W/\partial\lambda$, and of $\beta\partial^2 W/\partial\lambda^2$ across $\lambda = 3^{-1/2}$ provide three further conditions, which respectively can be shown from Eq. (III.13) to have the following explicit forms:

$$3^{-3/2} J + 3^{-1} K + 3^{-1/2} L + M + 3^{1/2} N$$

$$= -\ln\left\{1 - \frac{4\pi\rho}{3^{5/2}} + 2\pi^2\rho^2 \int_1^{2/3^{1/2}} \left[\frac{4r^2}{3^{5/2}} - \frac{r^3}{3} + \frac{r^5}{12}\right] g(r) dr\right\} \quad ; \tag{A.3}$$

$$J + 2\cdot 3^{-1/2} K + L - 3N$$

$$= \left\{1 - \frac{4\pi\rho}{3^{5/2}} + 2\pi^2\rho^2 \int_1^{2/3^{1/2}} \left[\frac{4r^2}{3^{5/2}} - \frac{r^3}{3} + \frac{r^5}{12}\right] g(r) dr\right\}^{-1}$$

$$\times \left\{\frac{4\pi\rho}{3} - 4\pi^2\rho^2 \int_1^{2/3^{1/2}} \left[\frac{2r^2}{3} - \frac{r^3}{3^{1/2}}\right] g(r) dr\right\} \quad ; \tag{A.4}$$

$$2\cdot 3^{1/2} J + 2K + 2\cdot 3^{3/2} N$$

$$= \left\{1 - \frac{4\pi\rho}{3^{5/2}} + 2\pi^2\rho^2 \int_1^{2/3^{1/2}} \left[\frac{4r^2}{3^{5/2}} - \frac{r^3}{3} + \frac{r^5}{12}\right] g(r) dr\right\}^{-2}$$

$$\times \left\{\frac{4\pi\rho}{3} - 4\pi^2\rho^2 \int_1^{2/3^{1/2}} \left[\frac{2r^2}{3} - \frac{r^3}{3^{1/2}}\right] g(r) dr\right\}^2$$

$$\tag{A.5}$$

$$+ \left\{ 1 - \frac{4\pi\rho}{3^{5/2}} + 2\pi^2 \rho^2 \int_1^{2/3^{1/2}} \left[ \frac{4r^2}{3^{5/2}} - \frac{r^3}{3} + \frac{r^5}{12} \right] g(r) dr \right\}^{-1}$$

$$\times \left\{ \frac{8\pi\rho}{3^{1/2}} - 4\pi^2 \rho^2 \int_1^{2/3^{1/2}} \left[ \frac{4r^2}{3^{1/2}} - r^3 \right] g(r) dr \right\} .$$

These Eqs. (A.1)-(A.5) can be simultaneously solved for the five quantities $J$, $K$, $L$, $M$, and $N$.

Three additional equations emerge from the requirements that $\beta W_2$ and its first two derivatives with respect to $t$ (at constant $s$) be continuous across the locus $t = 0$, i.e. $\lambda = 1/2$. These equations are necessary to determine the remaining unknowns $X(s)$, $Y(s)$, and $Z(s)$. With the understanding that subscripts $i = 0,1,2$ respectively refer to continuity of $\beta W_2$, its first $t$ derivative, and its second $t$ derivative at $t = 0$, these three equations may be symbolically be written as follows:

$$A_i(s) = B_i(s, t = 0) \qquad (i = 0,1,2) . \qquad (A.6)$$

Here the $A_i(s)$ stem from the assumed asymptotic formula in Eq. (IV.5), while the $B_i(s,t)$ arise from Eqs.(III.11) and (III.12). The specific forms for the $A_i(s)$ are:

$$A_0(s) = \left[ \frac{1}{4} - \frac{3s^2}{16} - \frac{s^3}{16} \right] J + \left[ \frac{1}{2} + \frac{s}{4} \right] K + \frac{X(s)}{2} + Y(s) + 2Z(s) \qquad (s \leq 0) ,$$

(A.7)

$$= \frac{J}{4} + \frac{K}{2} + \frac{X(s)}{2} + Y(s) + 2Z(s) \qquad (0 \leq s) ;$$

$$A_1(s) = \left[ \frac{3}{4} - \frac{3s^2}{16} \right] J + \left[ 1 + \frac{s}{4} \right] K + \frac{X(s)}{2} - 2Z(s) \qquad (s \leq 0) ,$$

(A.8)

$$= \frac{3J}{4} + K + \frac{X(s)}{2} - 2Z(s) \qquad (0 \leq s) ;$$

$$A_2(s) = \frac{3J}{2} + K + 4Z(s) \qquad \text{(all } s\text{)} . \qquad (A.9)$$

One also has the following expression:

$$B_0(s,t) = -\ln\left\{ 1 - \pi\rho\left[ \frac{(t+1)^3}{3} - \frac{s^2(t+1)}{4} - \frac{s^3}{12} \right] + \frac{2\pi^2\rho^2}{(s+t+1)} \int_1^{s+2(t+1)} k(s - r_1 + t + 1, \frac{t+1}{2}) r_1 g(r_1) dr_1 \right\}$$

$$(s \leq 0) ,$$
(A.10)

$$= -\ln\left\{ 1 - \frac{\pi\rho(t+1)^3}{3} + \frac{2\pi^2\rho^2}{(s+t+1)} \int_{\max(s,1)}^{s+2(t+1)} k(s - r_1 + t + 1, \frac{t+1}{2}) r_1 g(r_1) dr_1 \right\} \qquad (0 \leq s) .$$

Here the integral kernel $k$ has been defined earlier in Eq. (III.12). One also has:

$$B_1(s,t) = \partial B_0(s,t)/\partial t \qquad (A.11)$$

and

$$B_2(s,t) = \partial^2 B_0(s,t)/\partial t^2 \, , \qquad (A.12)$$

the explicit forms of which are straightforward to generate but are rather lengthy, and so will not be shown here.

## Acknowledgments

P.G.D. gratefully acknowledges financial support by the U.S. Department of Energy, Division of Chemical Sciences, Geosciences, and Biosciences, Office of Basic Energy Sciences, grant no. DE-FG02-87ER13714.

## Tables

Table 1. Intervals of the size parameter $\lambda$ corresponding to the maximum numbers $n^*(\lambda)$ of unit-diameter spheres whose centers can fit simultaneously and without overlap within a spherical region of radius $\lambda$.

| $n^*(\lambda)$ | $\lambda$ interval |
|---|---|
| 1 | $0 < \lambda/\sigma < 1/2$ |
| 2 | $1/2 \leq \lambda/\sigma < 3^{-1/2} \cong 0.577350$ |
| 3 | $3^{-1/2} \leq \lambda/\sigma < (3/8)^{1/2} \cong 0.612372$ |
| 4 | $(3/8)^{1/2} \leq \lambda/\sigma < 2^{-1/2} \cong 0.707107$ |
| 5 | (no interval)[a] |
| 6 | $2^{-1/2} \leq \lambda/\sigma < 0.795627$ [b] |
| 7 | $0.795627 \leq \lambda/\sigma < (1/2)(2 + 2^{-1/2})^{1/2} \cong 0.822664$ |

[a] When $\lambda$ becomes large enough to allow 5 sphere centers, 6 can also fit within the spherical region.
[b] Reference [19].

**Figure Captions**

1. Double exclusion cavities considered in the present extension of scaled particle theory. The two spherical exclusion zones have a common variable radius $\lambda$. The distance $r \geq 0$ between their centers can produce overlap (top) or disconnection (bottom).
2. Extremal geometry for three sphere centers just to fit on the boundary of a ($r, \lambda$) double cavity. The three spheres are in mutual contact, and are coplanar with the centers of the two $\lambda$-spheres comprising the double cavity.
3. Regions in the $r, \lambda$ positive quadrant for $n_{\max} = 1, 2, 3$ and $\geq 4$.
4. Typical placement of three unit spheres inside a single radius-$\lambda$ sphere for $\lambda$ slightly in excess of the threshold value $3^{-1/2}$. The centers of the three unit spheres (1,2, and 3) are depicted as though those spheres are in mutual contact.
5. Relevant regions and linear loci in the positive $r, \lambda$ positive quadrant. Shown explicitly are the double-cavity contact line, and the horizontal locus that is relevant to evaluation of $g(r, \rho)$.

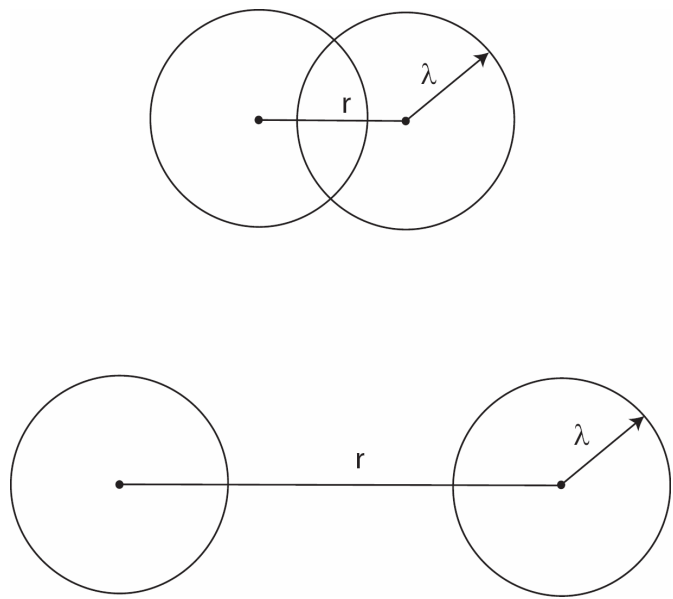

Figure 1

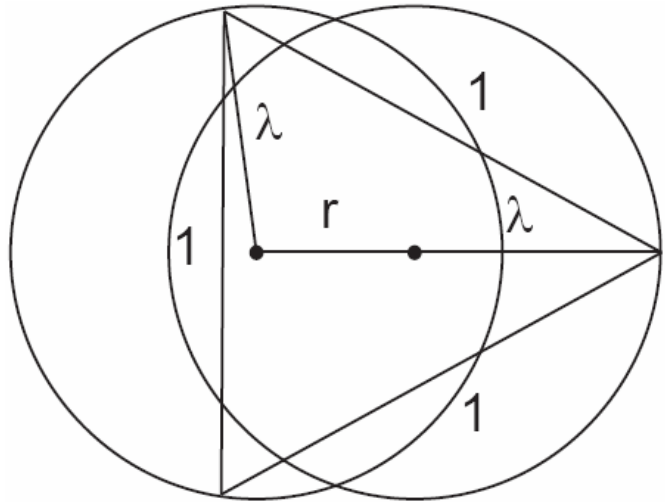

Figure 2

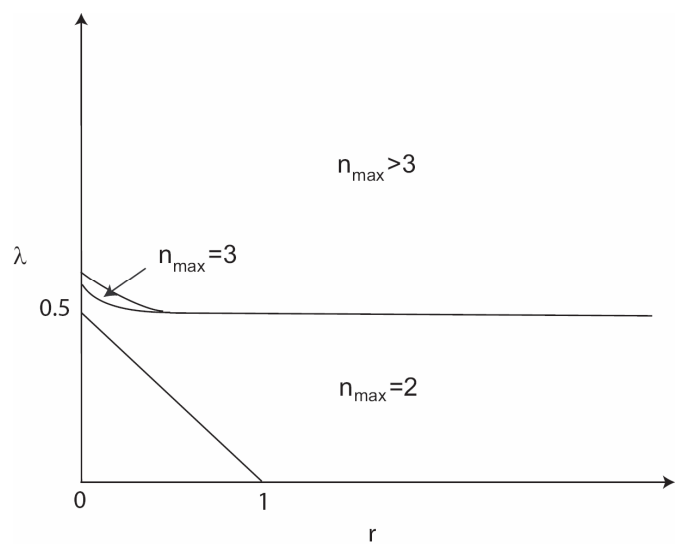

Figure 3

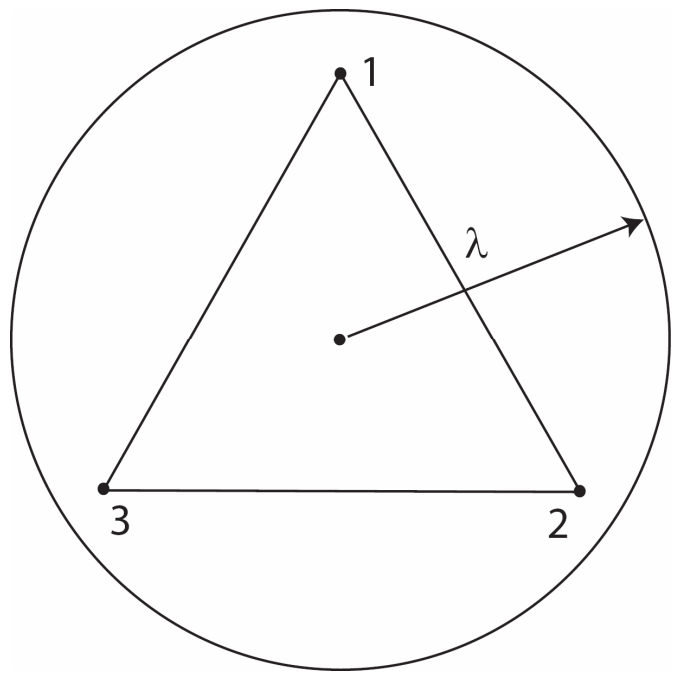

Figure 4

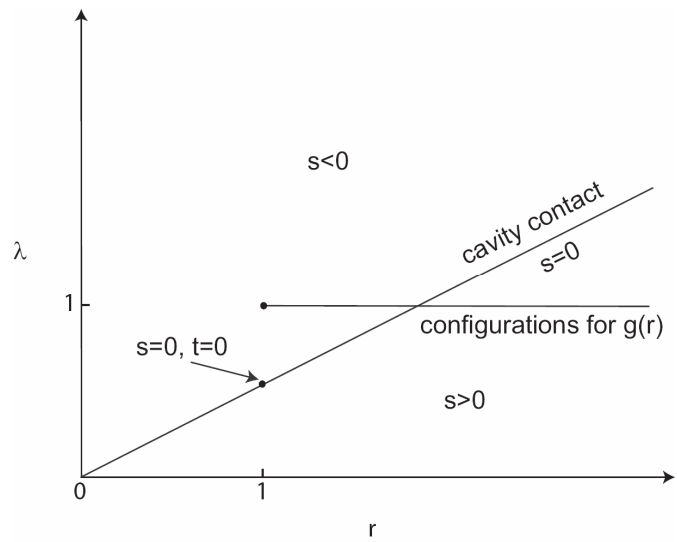

Figure 5